\documentclass[11pt]{amsart}
\usepackage{amscd,amsthm,amssymb,amsfonts,amsmath}
\newfont{\frakturfont}{eufm10 scaled\magstep1}

\newcommand{\frakj}{\mbox{\frakturfont j}}
\newcommand{\G}{\Gamma}
\newcommand{\g}{\gamma}

\newcommand{\db}{\bar{\partial}}

\newtheorem{theorem}{Theorem}[section]

\newtheorem{remark}[theorem]{Remark}

\newtheorem{define}[theorem]{Definition}

\title{Mechanical D branes and B fields}
\author{Mark Stern}
\date{}
\begin{document}

\begin{abstract}We represent $B$ fields and higher p-form potentials on a manifold $M$ as connections on affine bundles over $M$. We realize D branes on $M$ as special submanifolds of these affine bundles. We check the physical relevance of this representation by showing that the properties of the resulting branes closely correspond to those of their string theory counterparts. As a simple application of this geometric understanding of the $B$ field, we show how to obtain certain supersymmetry algebra deformations induced by 2 and 3 form potentials.
 \end{abstract} 
 \maketitle
\section{Introduction}
In this note we represent $B$ fields and higher $p-$ form potentials on a manifold $M$ as connections on affine bundles over $M$. We find that a $p-$form potential arises as a connection on an affine bundle locally modelled on either $\bigwedge^{p-1}T^*M$, discrete quotients of $\bigwedge^{p-1}T^*M$, or on $\bigwedge^{p}T^*M$. The representation on the locally $\bigwedge^{p-1}T^*M$ affine bundle requires the field strength of the potential to vanish in cohomology, the $\bigwedge^{p}T^*M$ representation leaves the field strength unrestricted, and the discrete quotient construction both restricts the field strength to lie in a discrete subgroup of the cohomology and restricts the geometry of $M$.

This simple geometric interpretation of $p-$ form potentials is inspired by string theory but not directly derived from it. Hence, we are forced to consider the question of whether this representation is physically germane. In order to address this question, we include $D$ branes and some rudimentary form of supersymmetry algebra into our model and study whether the relations between potentials, branes, and susy algebra corresponds to our string theory expectations. The representation of $D$-branes is fairly natural but the representation of the susy algebra is far from unique. This nonuniqueness is to be expected, given the variety of theories in which these objects appear, but restricts the definiteveness of some of our results. Even with this difficulty, our model strikes amazingly close to string theory in many ways.

For $B$ fields realized as connections on affine cotangent bundles, denoted $T_B^*M$, we model $D$ branes by holonomic submanifolds $Z$ of $T_B^*M$ with $dim(Z) = dim(M)$. Generically, these branes can be identified with submanifolds of $M$ equipped with a gauge field for a line bundle, or in the case of multiple parallel (in the fiber) branes, a gauge field for a higher rank vector bundle. To study the branes more closely, we assume that $M$ has an almost complex structure and construct an algebra generated by zero order and first order differential operators, analogous to an $N=2$ superconformal vertex algebra. The commutation relations for the supercharges impose the standard geometric constraints on $M$: integrability of the complex structure, compatibility of the complex structure and the metric, and equality of the torsion 3 form with $i(\partial -\db)$ of the Kahler form. The field strength of the $B$ field, however, satisfies a relation with the torsion, which is slightly twisted from the expected one. In order to implement this algebra without overly constraining the geometry of $M$, we are led to consider families of deformations of this algebra. These operator deformations can be interpreted as deformations of the affine structure of the bundle, replacing it with more complicated geometries, which to first approximation can range from the geometry of the neighborhood of the diagonal in $M\times M$ to classical analogs of noncommutative geometry.

We define a supersymmetry algebra on the brane $Z$ by orthogonally projecting the supercharges onto $Z$. For these models we find, as usual (see for example \cite{D1},\cite{OOY},  \cite{Sh}, and references therein), that imposing supersymmetry constraints forces $B$ type branes to correspond to holomorphic submanifolds equipped with a holomorphic bundle or more generally, a coherent sheaf. Moreover, we find possible geometric interpretations of D branes corresponding to exact sequences of coherent sheaves.  We also examine a $B$ brane cohomology associated to our $B$ brane data which we suspect may be related to "stringy cohomology". For $C^2/Z_2$, we argue (completely non rigorously) that the discrepancy between the dimensions of the usual cohomology and the $B$ brane cohomology is given by the dimension of the kernel of a superharmonic oscillator naturally arising when we attempt to deform one cohomology theory to the other, giving the same discrepancy in this single case as given by the stringy cohomology.

The $A$ branes in our model correspond to coisotropic (with respect to the almost Kahler form) submanifolds of $M$, equipped with a vector bundle whose curvature satisfies familiar constraints (see \cite{KOA}). When the brane $Y$ is Lagrangian, the curvature is twice the $B$ field (restricted to $Y$). In the higher dimensional coisotropic case, we find that the B field, curvature, and symplectic form restricted to $Y$ satisfy a slight twist of the relations found by Kapustin and Orlov (loc cit). The method of orthogonally projecting the supercharges onto $Z$ is not unique. We consider two natural procedures. One leads to the additional condition that the A-brane corresponds to a special Lagrangian submanifold. The second method imposes instead the condition that the brane correspond to a totally geodesic submanifold whose gauge field also determines a Killing vector field (possibly $0$).

At several points in the analysis of the representation of B fields by affine connections, we find that it is natural to consider $B$ fields varying in the fiber of the affine bundle. This leads to field strengths which are not naturally defined on $M$. In simple examples, Chern Simons terms must be added to the field strength in order to obtain a well defined 3 form on $M$. This suggests that the constructions here may also be useful for anomaly analysis.

The construction of $B$ fields as connections also shows immediately how these B fields deform supersymmetric Yang- Mills theories. Although these deformations can be shown to lead to the noncommutative supersymmetric Yang-Mills in the large $B$ limit (see \cite{SW}), they are more naturally given as the commutative formulation of this theory. Thus the realization of B fields as affine connections leads to a deformation which is analogous to the image of the Seiberg-Witten map (\cite{SW}).  It would be interesting to see what deformations are defined by the other $p$-form potentials. We examine briefly deformations of supersymmetry algebras associated to 3 forms. These turn out to be most tractable when one restricts attention to 6 dimensions and assumes the 3 forms are self-dual or anti-self dual. In this case, we show that the inductive procedure  for transforming the commutative formulation of the deformation to a noncommutative formulation (an "inverse Seiberg-Witten map") yields at first order a noncommutative product of one forms which is similar to the Poisson bracket. There is an obstruction to extending this noncommutative formulation to higher order.

Geometric realizations of B-fields from a different perspective have been considered in \cite{H1} and \cite{H2}. For a more abstract perspective see for example \cite{CJM} and references therein.

The remainder of this paper is organized as follows. In section 2, we construct the affine form bundles and show how p-form potentials may be realized as connections on them. In section 3, we define holonomic sections of the affine bundles and define D branes as holonomic submanifolds of the affine bundles. In section 4, we define the supersymmetry algebras we use and check that they impose standard geometric constraints on $M$. When they impose excessive constraints, we show how to interpret the extra constraints as defining modifications of the operators instead of constraining the geometry. In section 5, we define BPS branes and examine the structure of A and B type branes. In section 6, we discuss, in the context of examples, the role of singular branes in understanding the distinction between vector bundles and more general sheaves in B branes. We also speculate about realizing stringy cohomology as a B brane cohomology. Finally, in section 7, we use the geometric understanding of 2 and 3 form potentials to construct deformations of supersymmetry algebras.

\section{Connections on affine bundles}
In this section, we consider a geometric interpretation of the representation in
\cite{S1} of $B$ fields as connections on spaces of partial differential operators. We "dequantize" the partial differential operator representation and consider instead connections on affine bundles modelled on $T^*M$, for some Riemannian manifold $M$.

Let $\pi:T^*M\rightarrow M$ denote the natural projection map. Because a choice of
coordinates $(x^i)$ on $M$ induces a choice of coordinates $(x,p)$ on
$T^*M$, there is a coordinate dependent lift of the vectorfield
$\frac{\partial}{\partial x^i}$ (and therefore any vectorfield)
on the base to a vectorfield, also denoted $\frac{\partial}{\partial x^i}$, on $T^*M$.
The Riemannian metric and Levi Civita connection on $M$ induce a metric
on $T^*M$ and therefore allow us to define a coordinate independent lift
$$X\rightarrow X^h,$$
from tangent vectors on $M$ to tangent vectors on $T^*M$. Here $X^h$ is defined to be the unique vector on $T^*M$ satisfying $d\pi(X^h) = X$, and
$X^h$ is orthogonal to the tangent space of the fibers of $T^*M$.
In local coordinates, $X^h$ is given by

 $$X^h = X -
 p_j(\nabla_Xdx^j)(\frac{\partial}{\partial x^s})\frac{\partial}{\partial
p_s}.$$
In particular,
$$(\frac{\partial}{\partial x^i})^h = \frac{\partial}{\partial x^i} + p_n\Gamma^n_{is}\frac{\partial}{\partial p_s}.$$
Here we are using the same symbol to denote a vector field on the base and
its coordinate determined lift to $T^*M$.
We note that the metric on $T^*M$ is given by
$$g_{ij}dx^i\otimes dx^j
 + g^{ij}(dp_i - p_m\Gamma_{ik}^mdx^k)\otimes (dp_j - p_n\Gamma_{jp}^ndx^p).$$

We now wish to relax the vector bundle structure of $T^*M$ to an affine structure. Hence, we introduce new coordinate transformations acting as translations in the fiber. These transformations are of the form $(x,p)\rightarrow (y,q)$ where $y=x$ and
$$q= p+\lambda,$$
for a locally defined $1$ form $\lambda$.
Comparing the coordinate defined lifts in these coordinate systems, we see that
$$\frac{\partial}{\partial x^i} = \frac{\partial}{\partial y^i} + \lambda_{s,i}\frac{\partial}{\partial q_s}.$$
We may also rewrite this as
$$\frac{\partial}{\partial x^i} + p_n\G^n_{is}\frac{\partial}{\partial p_s} = \frac{\partial}{\partial y^i} + q_n\G^n_{is}\frac{\partial}{\partial q_s} + \lambda_{s;i}\frac{\partial}{\partial q_s}.$$
Thus the horizontal lift is no longer well defined independent of coordinate system.
To correct this, fix a locally defined 2 form $b$ and 1 form $\mu$ on $M$.
Define a lift of a vector field $X$ to a vector field $X^{b,\mu}$   as follows. In a coordinate neighborhood, define

$$H_i:=(\frac{\partial}{\partial x^i})^{b,\mu}
 :=  \frac{\partial}{\partial x^i} + (\mu_{i,k} +
 b_{ik} + (p_n-\mu_n)\Gamma^n_{ik})\frac{\partial}{\partial p_k}.$$

This extends linearly to a lift $X^{b,\mu}$ (or simply $X^h$ when $(b,\mu)$ is understood) of an arbitrary vector field $X$,
which is clearly independent of coordinate system $(x)$ on $M$.

Consider now an affine change of coordinates $y=x$ and
$$q_k = p_k + \lambda_k(x).$$
Then $H_k$ transforms as
$$(\frac{\partial}{\partial x^k})^{b,\mu} = (\frac{\partial}{\partial y^k})^{b+d\lambda,\mu+\lambda}.$$
Thus, if we take $(b,\mu)$ to be affine coordinate dependent, transforming as
$$(b,\mu)\rightarrow (b+d\lambda,\mu+\lambda),$$
then the lifts $H_i$ are well defined independent of the coordinate system.
 We denote such an affine bundle determined by the local data $B=(b,\mu)$ by
$$T_B^*M.$$
 We interpret the $H_i$ as horizontal lifts. Hence the pair $(b,\mu)$ becomes part of the metric data on $T_B^*M$.

Note that the canonical symplectic form $dx^k\wedge dp_k$ is no longer well
defined if we allow $p_k$ to vary in this fashion, as
$$dx^k\wedge dq_k = dx^k\wedge dp_k - d\lambda.$$
 The natural extension of the symplectic form
to this context is given by
$$\omega_B := dx^k\wedge (dp_k - V_{ik}dx^i),$$
where
$$V_{ik} := \mu_{i,k} + b_{ik} + (p_n-\mu_n)\Gamma^n_{ik}.$$

These structures are related to connections on line bundles.
Let $L$ be a line bundle with transition functions given by the
$g^{\alpha\beta}$. Then any connection $D$ on $L$ can be written in local
coordinates on each coordinate domain $U_{\alpha}$ as
$$D = d+a_{\alpha},$$
where the connection one forms $a_{\alpha}$ satisfy the usual transformation
 rules 
$$a_{\beta} = a_{\alpha} + \lambda^{\alpha\beta},$$
with
$$\lambda^{\alpha\beta} = (g^{\alpha\beta})^{-1}dg^{\alpha\beta}.$$
Hence, the connection forms are not sections of a vector bundle but of an
 affine bundle. A $B$ field is the additional data of a connection on this affine bundle
of connections.

Observe that the $\lambda^{\alpha\beta}$ satisfy the cocycle relation
\begin{equation}\label{cc1}
\lambda^{\alpha\beta} + \lambda^{\beta\nu} + \lambda^{\nu\alpha} = 0.
\end{equation}
As discussed in \cite[Section 11]{S1}, this cocycle condition implies that the
two form gauge field $b$ must have
cohomologically trivial flux $H = db$. In order to treat the case of 
nontrivial $H$, we must relax the cocycle condition.
\subsection{Discrete quotients}\label{discq}
In order to realize a relaxed cocycle structure geometrically, we
replace our local model $T^*U_{\alpha}$ by equivalence classes of 1 forms. In 
\cite{S1}, we interpreted this to mean that we must replace connections by gauge
equivalence classes of connections. However, less drastic equivalence classes may
be considered if the base manifold has more structure. For example, suppose that $T^*M$
 contains a {\em discrete} subbundle $L$ whose fibers $L_x$ are discrete abelian subgroups of $T_x^*M$. We assume that $L$ is compatible with the connection. In particular, it spans a flat vector subbundle. Then we consider an affine bundle locally modelled on the quotient bundle $T^*M/L$. The cocycle condition becomes
\begin{equation}
\lambda^{\alpha\beta}(x) + \lambda^{\beta\nu}(x) + \lambda^{\nu\alpha}(x) \in L_x,
\end{equation}
and $H$ need not be cohomologically trivial.

 The preceding construction requires that $M$ locally have a Euclidean factor and is thus suitable, for example, if $M$ is of the form $M=T^k\times N$. It is important to modify this construction so that we can consider quotients without passing to full gauge equivalence classes and without imposing such strong assumptions on the local geometry. For example, suppose we assume only that $M$ has a nontrivial Killing vector with metrically dual one form $K$, and consider quotients of $T^*M$ by the additive subgroup generated by $K$. Let $\Phi$ denote the map of $T^*M$ given by $(x,p)\rightarrow (x,p+K).$ Then
$$d\Phi_p(H_i) = H_i(p+K_idx^i) + \frac{1}{2}(dK)_{is}\frac{\partial}{\partial p_s}.$$
  Thus we cannot simultaneously impose the desirable conditions that $\Phi$ is an isometry and that the projection map is an isometry when restricted to the horizontal subspace. A simple modification of our construction which removes this conflict is to remove the assumption that the $B$ field is locally the lift of a 2 form on $M$ and to assume instead that it has the form
$$\tilde b(x,p) = b_0(x) + \frac{<p-\mu,K>}{2|K|^2}dK.$$
Now $d\tilde b$ is globally well defined on $T_B^*M$, but does not descend to a form on $M$.
Assume for simplicity that $|K| = 1$ is constant and choose local coordinates so that $K^i\frac{\partial}{\partial x^i} = \frac{\partial}{\partial x^0}$. Then
$$d\tilde b(x,p) = db_0(x) + (dp_0-d\mu_0)\wedge dK/2.$$
If we use the metric to assign a form on $M$ to the field strength via the map
$$dp_i-d\mu_i\rightarrow g_{ij}dx^j,$$
then $dp_0\rightarrow K$ and
$$d\tilde b(x,p) \rightarrow  db_0(x) + K\wedge dK/2.$$
Hence, we see that passing to discrete quotients forced us to introduce $B$ fields varying in the affine fiber. This in turn led to anomalous field strengths which can be interpreted as field strengths on $M$ by the addition of a Chern-Simons term to the original field strength.

\subsection{Gauge equivalence classes}
Following \cite{S1}, we next consider quotienting by gauge equivalence class. This makes sense as sheaves of
sections but not as bundles, since the relation $A\sim A+g^{-1}dg$ is not
defined pointwise on open sets $U_{\alpha}$. To remedy this, replace the local model
$T^*U_{\alpha}$ by its first jet space, whose fiber at any $x\in U_{\alpha}$
  consists of pairs $(a,\phi),$ where $a\in T_x^*U$ and 
  $\phi\in Hom(T_xU_{\alpha},T_aT_x^*U_{\alpha}).$
We use local coordinates $(p_k,\tilde p_{ji})$ to denote a point in the fiber.  
The $1$ jet associated to a section $A$ of $T^*U_{\alpha}$, with $A = p_kdx^k$
in local coordinates is then given by $\tilde p_{ji} = \frac{\partial
p_j}{\partial x^i}.$
To approximate the quotient of one forms by exact forms to first
order, we drop $a$ and replace the two tensor $\tilde p_{ji}dx^i\otimes dx^j$ by 
the two form $p_{ji}dx^j\wedge dx^i$, where 
$p_{ji} = (\tilde p_{ji} - \tilde p_{ij})/2$.
Our jet space approximating gauge equivalence classes of connections then becomes the bundle of two forms.

A section of a jet space is
called holonomic if it is the jet of a smooth section of the original 
bundle. As our two forms were jets of 1 forms mod exact forms, we should call 
a section holonomic if it is an exact 2 form and locally holonomic
if it is closed. Viewing $T^*M$ as the space of 1 jets of functions, then its
sections are holonomic if they are given by exact 1 forms and locally holonomic if
they are given by closed 1 forms. Equip $T^*M$ with its usual symplectic form, $dx^i\wedge dp_i$. Recall that a section of $T^*M$ is a Lagrangian submanifold if and only if it is given by a closed 1 form. Similarly, it is easy to see that a section of
$\bigwedge^2T^*M$ is locally holonomic if and only if the pullback of
$W_2:=dx^i\wedge dx^j\wedge dp_{ij}$ to the section vanishes. Hence, $W_2$
may be viewed as the analog of the symplectic form on $T^*M$. We may define
similar forms $W_p:= dx^{i_1}\wedge\cdots\wedge dx^{i_p}dp_{i_1\cdots i_p}$ on 
$\bigwedge^pT^*M$.

We consider lifts of vectorfields $X$ from $M$ to $\bigwedge^2T^*M$. 
The horizontal lift $X^h$ of $X$ is given by 
$$X^h = dx^i(X)\frac{\partial}{\partial x^i} + 
dx^i(X)(p_{pk}\Gamma_{ij}^p +
p_{jp}\Gamma_{ik}^p)\frac{\partial}{\partial p_{jk}}.$$

Suppose we wish to allow affine changes of coordinates in this bundle 
$$p_{jk}\rightarrow p_{jk} + \lambda_{jk},$$
where $\lambda$ is a two form. Then we need to introduce a local 3 form $c$ and a local two form $b$ and define the lift
$$X^{C,b} = dx^i(X)\frac{\partial}{\partial x^i}
+ dx^i(X)(b_{ji,k}+b_{ik,j} + c_{ijk} + (p-b)_{pk}\Gamma_{ij}^p +
(p-b)_{jp}\Gamma_{ik}^p)\frac{\partial}{\partial p_{jk}}.$$
Then this lift is well defined if when $p\rightarrow p+\lambda$,
$$b\rightarrow b-\lambda,$$
and
$$c\rightarrow c+d\lambda.$$
We denote an affine bundle determined by the data $C=(b,c)$ by $\bigwedge^2T^*_C M.$
For unrestricted gauge parameters $\lambda$, this gives a geometric realization of 3 form potentials.
On the other hand, if we restrict to bundles with transition parameters given by
exact $\lambda = dv$, then we may take $c =0$ and locally have the transformation rule $b\rightarrow b+dv$. This yields a geometric realization of a 2 form potential with no restrictions on its field strength or the geometry of $M$.

The extension of $W_2$ to the affinized 2 form bundle is now given by
$$W_{2,C} = dx^i\wedge dx^j\wedge (dp_{ij} - V_{kij}dx^k),$$
where
$$V_{ijk} = b_{ji,k}+b_{ik,j} + c_{ijk} + (p-b)_{pk}\Gamma_{ij}^p +
(p-b)_{jp}\Gamma_{ik}^p.$$

These constructions (either unrestricted or exact gauge parameters) are readily generalized to obtain $p-$ form potentials for arbitrary $p$.

\section{Mechanical D branes}\label{MDB}

In the presence of a $B$ field, we must modify our definition of a Lagrangian or holonomic section.
\begin{define}Let $\Lambda_2$ be a $2-$ form on $T_B^*M$ which is the pullback of a $2-$ form, also denoted $\Lambda_2$, on $M$. A submanifold $Z$ of $T_B^*M$ is $\Lambda_2$-Lagrangian or $\Lambda_2$-holonomic if the pull back of $\omega_B-\Lambda_2$ to $Z$ is $0$. Given a 3 -form, $\Lambda_3$, on $\bigwedge^2T^*_C M$ which is the pullback of a $3-$ form on $M$, we say a submanifold $Z_2$ of $\bigwedge^2T^*_C M$ is $\Lambda_3-$holonomic if the pull back of $W_{2,C}-\Lambda_3$ to $Z_2$ vanishes. Similarly, we may define $\Lambda_p$-holonomic submanifolds of affine p-form bundles. We will frequently omit the prefix $\Lambda_p$ when $\Lambda$ is understood.
\end{define}

We propose that, given a choice of $\Lambda$, $D$ branes in our model should be given by $n=$ $dim M$ dimensional $\Lambda-$ holonomic submanifolds. Let $X^i(s,t)$ denote coordinate functions of a string mapping into $M$. Here $t$ denotes a timelike coordinate. Under the rough correspondence
$$g_{ij}X^j_t\rightarrow H_i, \mbox{      } X^i_s\rightarrow \frac{\partial}{\partial p_i},$$
a choice of holonomic submanifold corresponds to a choice of boundary conditions for the $X^i$, at least when $\Lambda = 0$.  (In fact, our original definition had $\Lambda=0$, but later computations suggested that this was too tight a constraint and that perhaps D branes (not necessarily BPS) should be defined by an open condition, as is given by introducing arbitrary $\Lambda$.) We will focus primarily on the case of $T_B^*M$. (No quotients imposed unless otherwise stated). Observe that the modified symplectic form $\omega_B$
is nondegenerate but need not be closed:
$$d\omega_B = d\hat V,$$
where $\hat V := V_{ik}dx^i\wedge dx^k$. When $\Gamma$ is torsion free then in $\mu = 0$ local coordinates,
$\hat V = b$, and we get
$$d\omega_B = \hat H,$$
where $\hat H:= db$ is the field strength of $b$.
 A local section, $\alpha(x) = p_k(x)dx^k$, of $T_B^*M$, is $\Lambda$ Lagrangian if and only if
$$d\alpha = \hat V + \Lambda.$$
Thus if $\Gamma$ is torsion free and $\Lambda = 0$,  we see that there are no Lagrangian sections unless, in $\mu = 0$ coordinates, $\hat H=0$. (There is, however, no topological  obstruction to an affine bundle admitting a continuous section). In flat space with a constant torsion tensor and $\Lambda = 0$, the condition that the radial brane given by the section $dr^2$ is Lagrangian is the familiar
$$\hat H = - g_{nm}dx^n\G^m_{ik}dx^i\wedge dx^k.$$

All the branes we are considering are  n dimensional
submanifolds of $T_B^*M^n$. The designation of $p-$
brane, however, should be reserved for those Lagrangian submanifolds
whose projection onto $M^n$ is $p$ dimensional.

In order to make detailed computations, we restrict ourselves
to a large (locally generic) class of branes $Z$. For $Z$ connected, we assume that
we may choose local coordinates so that in each coordinate neighborhood, $Z$ is
(the intersection with the coordinate neighborhood) of the conormal
bundle of a submanifold $S$ of $M$. Under an affine change of coordinates in the
fibers, this local structure is replaced by the conormal bundle of a section of
$T_B^*M$ over the submanifold $S$ of $M$. Here the notion of "conormal bundle of
a section" is defined by the equality of the two structures. More generally, we allow $Z$ to have components of the above form.

Once we have fixed affine coordinates identifying a component of $Z$ with the conormal bundle of a submanifold $S$, we no longer have the freedom to set $\mu_i$ to zero for $\frac{\partial}{\partial x^i}$ tangent to $S$. Hence $S$ comes equipped with a 1 form whose exterior derivative is constrained by the $\Lambda$-holonomic condition to be twice $b-\Lambda$ restricted to $S$; in other words, a gauge field for a line bundle with local frame. If $Z$ has $k$ components all of the form of a conormal bundle over a fixed manifold $S$, then we may view $S$ as equipped with a gauge field for a sum of $k$ line bundles. This data is not sufficient, however, to determine $Z$. The relative positions of the components must also be fixed. This data is given by globally defined one forms (not gauge).  Hence the data is more constrained than that of an arbitrary connection on a rank $k$ bundle.

Consider next a section $f= p_{jk}dx^j\wedge dx^k$ of $\bigwedge^2T^*M$. The pull back
by $f$ of $W_2$ is $df$ and as previously noted, $f$ is a  holonomic section of
$\bigwedge^2T^*M$ only if it is closed. In the presence of a $c$
field, however, the pullback of $W_{2,C}$ under a local section $f$ now vanishes if
$$df = dx^i\wedge dx^j\wedge dx^k V_{ijk}.$$
When $\Gamma$ is torsion free and $\Lambda = 0$, this gives
$$dx^i\wedge dx^j\wedge dx^k V_{ijk} = c - \frac{2}{3}db,$$
and $c$ must be closed. If $c$ is zero, then appropriate 
multiples of $b$ itself provide holonomic sections.
\section{A supersymmetry algebra}
\subsection{The operators}\label{ABC}
In this section we define 4 differential and 2 algebraic operators on sections of vector bundles over $T^*_BM$ which are analogous to the supercharges and the $U(1)$ generators of the N=2 superconformal vertex algebra associated to an almost complex manifold $M$ equipped with an almost complex structure $J$, metric $g$, associated 2 form $\omega$, and conformal factor $e^{\phi}$.
These operators arise in the following fashion. Let $X^j(z)$ denote the coordinates of a string worldsheet in $M$. Let $\psi$ and $\bar \psi$ be associated fermionic fields.
Following \cite[subsection 4.2]{KO} consider the supercharges corresponding to
$$Q^{\pm}(z) = \frac{i}{\surd 32}g(\psi(z),\partial X(z))\pm \frac{i}{\surd 32}\omega(\psi(z),\partial X(z)),$$
and
$$J(z) = \frac{-i}{2}\omega(\psi(z),\psi(z)).$$

Write $z=t+is$ with $t$ a timelike parameter and $s$ spacelike. In the low energy approximation where we replace a string with a point particle, $X^i_t$ enters the associated quantum mechanics as
$\nabla_{\frac{\partial}{\partial x^i}}$. Heuristically, the following constructions may be thought of as replacing $X^i(t,s)$ with its one jet in the s variable before passing to the point particle approximation. It would be interesting to examine the geometry of higher jets in the point particle approximation.  In \cite{S1}, we saw that in order to obtain a quantum mechanics on $M$ which behaved nicely under $T$ duality, it was necessary to associate to
$X_s^i$ an operator we denoted $ad(x^i)$ which acted as a derivation on partial differential operators. Then our quantum mechanics was defined on spaces of partial differential operators or on connections. To obtain a more geometric model, it is natural to replace $ad(x^i)$ by
$e^{\phi}g_{is}\nabla_{\frac{\partial}{\partial p_s}}$ and $\nabla_{\frac{\partial}{\partial x^i}}$ by
$\nabla_{(\frac{\partial}{\partial x^i})^h}$. Using this dictionary, we now obtain simple differential operators modelled on the above charges.

Extend the almost complex structure operator $J$ to
$TT^*_{B}M$ by requiring it to commute with horizontal lift:
$$JH_i = J_i^kH_k,$$
where $J\frac{\partial}{\partial x^i} = J_i^k\frac{\partial}{\partial x^k}.$
The complex structure then has a natural extension to the vertical tangent
vectors given by the contragredient:
$$\tilde J\frac{\partial}{\partial p_i} = -J^i_k\frac{\partial}{\partial p_k}.$$
Following \cite{E}, we call this choice of complex structure on the vertical
tangent space the {\em geometric } complex structure. In general one may wish to
consider an arbitrary extension of the complex structure to this
subspace.

 Let $\psi$ and $\bar \psi$ be 2 elements of 2 (possibly distinct)
 Clifford ($TM$) modules. Let $\gamma^j\psi$ and $\gamma^j\bar\psi$ denote Clifford
 multiplication of $\psi$ by by $dx^j$.  Consider a new module admitting a Clifford
 representation of these sections, $\g^j\psi$ and $\g^j\bar\psi$. Thus we assume that $\gamma^j\psi$
  and $\gamma^j\bar\psi$ act as endomorphisms of this module and these
  endomorphisms satisfy the Clifford relations 
  $$\{\g^j\psi,\g^k\psi\} = 2g^{jk}|\psi|^2,$$
  and similarly for $\bar\psi$. 
   We assume that the the module is equipped with a compatible connection $\nabla$. In particular, we assume that
  $\nabla$ satisfies the Leibniz rule with respect to Clifford multiplication.

Fix a smooth function $\phi$ and set
$$\tilde g:= e^{\phi}g.$$
 Define operators
$$q^+ = (\g^i+iJ\g^i)\psi(\nabla_{H_i+iJH_i}
 + \nabla_{(\tilde g_{ij}+i(\tilde g\tilde J)_{ij})\frac{\partial}{\partial p_j}})/2,$$

$$q^- = (\g^i-iJ\g^i)\psi(\nabla_{H_i-iJH_i}
 + \nabla_{(\tilde g_{ij}-i(\tilde g\tilde J)_{ij})\frac{\partial}{\partial p_j}})/2,$$

$$\bar q^+ = (\g^i+iJ\g^i)\bar\psi(\nabla_{H_i+iJH_i}
 - \nabla_{(\tilde g_{ij}+i(\tilde g\tilde J)_{ij})\frac{\partial}{\partial p_j}})/2,$$

$$\bar q^- = (\g^i-iJ\g^i)\bar\psi(\nabla_{H_i-iJH_i}
 - \nabla_{(\tilde g_{ij}-i(\tilde g\tilde J)_{ij})\frac{\partial}{\partial p_j}})/2,$$

$$q = \g^i\psi(\nabla_{H_i} + \tilde g_{ij}\nabla_{\frac{\partial}{\partial p_j}}),$$

$$\bar q = \g^i\bar\psi(\nabla_{H_i}
 - \tilde g_{ij}\nabla_{\frac{\partial}{\partial p_j}}).$$

Here $J\g^i:=J^i_j\g^j.$

We also define the $U(1)$ generators
$$\frakj = (i/2)g_{lj}J\g^l\psi\g^j\psi.$$
$$\bar \frakj = (i/2)g_{lj}J\g^l\bar\psi\g^j\bar\psi.$$

In analogy with the $N=2$ superconformal vertex algebra, we require these operators to satisfy
$$0 = (q^+)^2 = (q^-)^2 = (\bar q^+)^2 = (\bar q^-)^2,$$
$$[\frakj,q^{\pm}] = \pm q^{\pm},$$
$$[\bar\frakj,\bar q^{\pm}] = \pm \bar q^{\pm},$$
and
\begin{equation}\label{mixedmode}
0 = \{q^+,\bar q^{\pm}\} = \{q^-,\bar q^{\pm}\}.
\end{equation}
\begin{remark}\label{escape} A natural modification of this construction is to  relax (\ref{mixedmode}) by including a central charge on the left hand side of that equation.
\end{remark}
\begin{remark}\label{dilo} Our choice of placement of the conformal factor $e^{\phi}$ is motivated in the following way. The natural operator to associate to the 3 form potential theory (which we will discuss later) is of the form
$$\g^j\psi\nabla_{(\frac{\partial}{\partial x^j})^h} + \g_{jk}\psi\nabla_{\frac{\partial}{\partial p_{jk}}}.$$
Upon dimensional reduction from say $11$ dimensions, this yields (among other things) the operator
$$\g^j\psi\nabla_{(\frac{\partial}{\partial x^j})^h} + \g_{j11}\psi\nabla_{\frac{\partial}{\partial p_{j11}}}.$$
The metric term $g_{11 11}$ now introduces a conformal factor into the last term, and $\psi$ which are $\pm 1$ eigenvectors of $\g^{11}$ lead to barred and unbarred $\psi$ in the dimensionally reduced theory.
\end{remark}
\subsection{The $U(1)$ charges}
In the (worldsheet) superconformal vertex algebra which we are trying to imitate 
with this target space operator algebra, the anticommutator of the
supercharges $Q^{+}$ and $Q^{-}$  is the sum of a term analogous to a Laplacian 
and a $U(1)$ current corresponding to $\frakj$ (see for example
\cite{KO}). Our target space analog $q^{\pm}$ of the supercharges are
degenerate generalized Dirac operators; hence their anticommutator is given by a nonelliptic analog of the rough
Laplacian plus a curvature term. In order to mirror the superconformal vertex 
algebra, $\frakj$ should appear as a summand of the curvature term. The natural way to
incorporate this is to assume that the $q$ act on sections of the tensor product of a
Clifford module with  a line bundle ${\mathcal L}$. The affine cotangent bundle 
may be thought of as an approximate moduli space for connections on ${\mathcal 
L}$. Hence the canonical connection for ${\mathcal L}$ pulled back to $T^*_BM$ 
would be
$$\frac{\partial}{\partial x^s} + i(p_s-\mu_s).$$
 This does not quite yield the correct commutation relation. In order to obtain 
the desired algebra, we replace this connection with
 $$\frac{\partial}{\partial x^s} + iae^{-\phi} J_s^t(p_t-\mu_t),$$
 for some real scalar $a$.

 This introduces a curvature term into
 $q^2$, given in $\mu = 0$ coordinates by
 $$iae^{-\phi}[\g^i\psi,\g^j\psi]((V_{iJj} +\tilde g_{Jji})
  + J_{j,i}^tp_t - \phi_iJ_{j}^tp_t),$$
 which is proportional to $\frakj$ in the special case when the $B$ and $\phi$ fields are trivial, $J$ integrable, and the connection Kahler. Here we have introduced the notation
 for tensors $L$,
 $$L_{iJk}:= L_{ij}J^j_k.$$
  We will show in the next section that $(q^+)^2 = 0$ implies that $J$ is integrable; so, we will assume this for the remainder of this subsection and compute in coordinates in which the $J_s^t$ are locally constant and $\mu = 0$.

 When the $B$ field is nontrivial, this computation suggests that we should
 either modify $g_{iJk}$ by the skew symmetric part of
 $$a(b_{iJk} + p_n(\Gamma^n_{iJk} - J_{k}^n\phi_i)),$$
 or modify $b$, allowing it to vary in the fiber,  so that this expression vanishes.
 In this note we will not analyze the case with $\frakj$ varying in the
 fibers and will postpone further consideration of $b$ varying in the fiber. Hence, we will assume that
 \begin{equation}\label{jtor}
 0 = \Gamma_{iJj}^n  - \Gamma_{jJi}^n - J_{j}^n\phi_i + J_{i}^n\phi_j.
 \end{equation}
 When $b_{ik}dx^i\wedge dx^k$ is of Hodge type $(1,1)$, the skew
symmetric part of $b_{Jik}$ vanishes. Hence, we need not modify the Kahler form
$\omega$ unless $b$ has a nonzero $(0,2)+ (2,0)$ component.

 Let $b^{[2]}$ denote the  $(0,2) + (2,0)$ component of $b$. Then  the preceding
 computation suggests that (for $b^{[2]}$ small) we replace $\frakj$ by
$$\frakj_b := (-i/2)(g_{Jij} + b^{[2]}_{ij})\g^i\psi\g^j\psi.$$
We cannot interpret this simply as a modification of the metric, since $b$ is 
skew. Hence, we must view this as simultaneously modifying the metric and the
complex structure. Perhaps, for $b^{[2]}$ small, this leads to an
iterative procedure for creating a new metric and complex structure determined by
$b^{[2]}$. Alternatively, we might wish to alter the $U(1)$ connection term
$iae^{-\phi} J_s^t(p_t-\mu_t)$ to $iae^{-\phi} \hat J_s^t(p_t-\mu_t)$ for some modified tensor $\hat J$ satisfying
$$((b+g){\hat J ik} - (b+g){\hat J ki})/2 = g_{Jik}.$$
We will not attempt to analyze either of these possibilities here.
Hence for the remainder of this note we impose the assumption:
\begin{equation}\label{as1}
b^{[2]} = 0
\end{equation}
when analyzing systems with this symmetry algebra.
Observe that under this restriction, the condition that a section $\alpha$ be 
Lagrangian  with $\Lambda = 0$ implies
$$(d\alpha)^{[2]} = \alpha_nT^n,$$
where $T$ denotes the torsion tensor
$$T^n_{st} = (\G^n_{st} - \G^n_{ts})/2.$$
 In particular, in the torsion free case, a
Lagrangian section corresponds to a connection whose $(0,1)$ component determines
a holomorphic structure on a line bundle.
\begin{remark} The connection term, $iaJ(p-\mu)$, which we have added to our
operators is defined on affine bundles but not on their quotients.  Hence, when using this construction, we are restricted
to the case where the field strength of the B field is cohomologically
trivial.\end{remark}

\subsection{Consequences of supersymmetry}
The purpose of this section is to check the extent to which our (1 jet) target space supersymmetry algebra reproduces standard stringy and supergravity relations between the various fields and the geometry of $M$ and therefore might serve as a reasonable tool for probing our model of $B$ fields as connections on affine bundles. We show integrability of the almost complex structure and the expected relation between the torsion tensor and the Kahler form follow from our algebra. Because we have not yet narrowed down our choice of Clifford modules in which $\psi$ and $\bar \psi$ sit, we will leave many relations unexplored.

We examine first the condition that
$$[q^+,\frakj] = - q^+.$$
Let
$$E^i:= (\g^i+iJ\g^i)\psi,$$
$$e^i:= (\g^i+iJ\g^i)\bar\psi,$$
$$Z_i = H_i+iJH_i + (g_{ij}+i(J g)_{ij})\frac{\partial}{\partial p_j},$$
and
$$\hat Z_i = H_i+iJH_i - (g_{ij}+i(J g)_{ij})\frac{\partial}{\partial p_j}.$$
The commutator, $[q^+,\frakj],$  is given by
$$E^i[\nabla_{Z_i},\frakj]/2 +
 (1/2)[ig_{Jsj}g^{is}\g^j - iJ\g^i - J^i_tg_{Jsj}g^{ts}\g^j - \g^i]\psi|\psi|^2\nabla_{Z_i}.$$
From the coefficient of $\nabla_{Z_i}$ we see that $g$ must be
compatible with the complex structure:
$$g_{Jij} = -g_{iJj},$$
and
$|\psi|=1$.

Let $R$ denote the curvature of $\nabla$, and let $R^{\G}$ denote the curvature of the connection $\G$.
In this notation, the condition that $(q^+)^2 = 0$ can be written as 
$$0 = E^s\nabla_{Z_s}E^t\nabla_{Z_t} = $$
$$E^sE^tR(Z_s,Z_t)/2 + E^sE^t\nabla_{[Z_s,Z_t]}/2 +
\{q^+,E^t\}\nabla_{Z_t}.$$
Observe that 
$$E^s = [q^+,x^s]$$
and $(q^+)^2 = 0$ imply
\begin{equation}\label{exactx}
\{q^+,E^t\} = 0.
\end{equation}
Thus we obtain
$$0 = E^sE^tR(Z_s,Z_t),$$
and 
$$0 = [Z_s,Z_t].$$

We next compute this commutator. 
Observe first that 
$$[H_s,H_t] = (V_{tu,s}-V_{su,t} + V_{sv}\G^v_{tu} - 
V_{tv}\G^v_{su})\frac{\partial}{\partial p_u} = :r_{stu}\frac{\partial}{\partial
p_u},$$
and
$$[H_s,\frac{\partial}{\partial p_t}] = -\G_{su}^t\frac{\partial}{\partial 
p_u}.$$
This gives 

$$0 = [Z_s,Z_t] = $$
$$(r_{stu}+ir_{sJtu}- ir_{tJsu}-
r_{JsJtu}))\frac{\partial}{\partial p_u} +
$$
$$(\tilde g_{tu,s}+i\tilde g_{Jtu,s}-\tilde g_{su,t} -
 i\tilde g_{Jsu,t})\frac{\partial}{\partial p_u}  + $$
$$  + (J_{s,y}^vJ^y_t - J_{t,y}^vJ_s^y + iJ_{t,s}^v-iJ_{s,t}^v)H_v $$
$$+ (i(\tilde g_{tu,Js}+i\tilde g_{Jtu,Js})
 - i(\tilde g_{su,Jt}+i\tilde g_{Jsu,Jt}))\frac{\partial}{\partial p_u}
$$
$$+ ((\G_{tu}^j+i\G_{Jtu}^j)(\tilde g_{sj}+i\tilde g_{Jsj})
- (\G_{su}^j+i\G_{Jsu}^j)(\tilde g_{tj}+i\tilde g_{Jtj}))\frac{\partial}{\partial p_u}
.$$

The vanishing of the horizontal component yields

\begin{equation}\label{jint}
0 = J_{s,t}^u-J_{t,s}^u + J_{s,y}^vJ_v^uJ^y_t - J_{t,y}^vJ_v^uJ_s^y.
\end{equation}
This, of course, is simply the condition that $J$ is integrable on the base
space, which we assume henceforth. Thus we have obtained the usual condition that $M$ is a hermitian complex manifold. Integrability implies that we can choose
local complex coordinates $z^s$. In real coordinates which are the real and
imaginary part of complex coordinates, the $J_s^t$ are constant. We will
restrict ourselves to such coordinate systems. 

If we assume that $b,\mu,\G,g$ are all constant on the fibers, then the vanishing of the terms in $[Z_s,Z_t]$ homogeneous of degree $1$ in $p$,
implies
\begin{equation}\label{holot}
R^{\G}_{st} = R^{\G}_{(Js)(Jt)}.
\end{equation}
Hence $TM$ has a holomorphic structure.
The remaining vanishing condition becomes 

$$0 = r_{stu}(0)-r_{JsJtu}(0) + \tilde g_{tu,s} - \tilde g_{su,t} + \tilde g_{sv}\G^v_{tu} - \tilde g_{tv}\G^v_{su}$$
$$- \tilde g_{Jtu,Js} + \tilde g_{Jsu,Jt} - \tilde g_{Jsv}\G^v_{Jtu}
 + \tilde g_{jtv}\G^v_{Jsu},$$
or
\begin{equation}\label{ono}
0 = (b_{tu,s}+\tilde g_{tu,s}) - (b_{su,t}+\tilde g_{su,t}) + (b_{sv}+\tilde g_{sv})\G^v_{tu}
\end{equation}
$$ - (b_{tv}+\tilde g_{tv})\G^v_{su} - (b_{Jtu,Js}+\tilde g_{Jtu,Js}) +$$ $$(b_{Jsu,Jt}+\tilde g_{Jsu,Jt}) -
(b_{Jsv}+\tilde g_{Jsv})\G^v_{Jtu} + (b_{Jtv}+\tilde g_{Jtv})\G^v_{Jsu}.$$
 Cyclically symmetrizing in $s,t,$ and $u$ in (\ref{ono}) and using (\ref{jtor}) gives

\begin{equation}\label{2tor}
0 = db_{stu} + b_{uv}T_{st}^v + b_{sv}T_{tu}^v  + b_{tv}T_{us}^v + H_{stu} + i(\partial - \db)\omega_{sut},
\end{equation}
where
$$H_{stu} = \tilde g_{sv}T^v_{tu} + \tilde g_{tv}T^v_{us} + \tilde g_{uv}T^v_{st},$$
and $\omega_{st} = \tilde g_{Jst}$.
Expanding similarly $0 = [\hat Z_s,\hat Z_t]$ (and continuing to assume the horizontal lifts are the same for both barred and unbarred
operators) we obtain
\begin{equation}\label{holoaff}
r_{stu}(0) = r_{JsJtu}(0),
\end{equation}
 and
\begin{equation}\label{tort}
 0 = g_{tu,s} + g_{tu}\phi_{s} - g_{su,t}- g_{su}\phi_{t} + g_{sv}\G^v_{tu} - g_{tv}\G^v_{su}
 - g_{Jtu,Js}\end{equation}
 $$- g_{Jtu}\phi_{Js} + g_{Jsu,Jt}+ g_{Jsu}\phi_{Jt} - g_{Jsv}\G^v_{Jtu}
 + g_{Jtv}\G^v_{Jsu}.$$

Equations (\ref{holoaff}) and (\ref{holot}) should be interpreted as
a holomorphic structure on the affinized cotangent bundle. If we cyclically
symmetrize (\ref{tort}) in $t,u,s$, we obtain (using (\ref{jtor}))
\begin{equation}\label{tort2}
 i(\partial - \db)\omega = H
 ,\end{equation}
 where $\omega_{ik} = g_{Jik}$.
 Compare to equation (2.17) in \cite{Stro}.
 If we set
$$\tilde H_{stu} = b_{sv}T^v_{tu} + b_{tv}T^v_{us} + b_{uv}T^v_{st},$$
then we also obtain
$$db_{stu} = - \tilde H_{stu}.$$
This is similar but not identical to the usual relation (see \cite{Stro}): $db = -H$.

\subsection{Relating left and right movers}
In this section, we continue to study the consequences of our supersymmetry algebra by examining the relations between $q's$ and $\bar q's$. Here we obtain results which suggest modifications of our algebra are required. In particular, if we require $\{q^+,\bar q^{\pm}\} = 0$, then we find that $M$ is required to be flat. In the next subsection we explore a modification of our operators which removes this geometric restriction.

Computing the anticommutators $\{q^+,\bar q^{\pm}\}$, we find
$[Z_s,\hat Z_t]$ must lie in the span of the $Z's$ and $\hat Z's$, and
$[\bar Z_s,\hat Z_t]$ must lie in the span of the $\bar Z's$ and $\hat Z's$.
Computing these vector commutators, we have
$$[Z_s,\hat Z_t] = [Z_s+\hat Z_s,\hat Z_t] = $$
$$2(\tilde g_{tu,s} + \tilde g_{tJu,Js} - \tilde g_{tv}\G^v_{su} +
\tilde g_{Jtv}\G^v_{(Js)u})(\frac{\partial}{\partial p_u} -
iJ_y^u\frac{\partial}{\partial p_y}).$$
Thus we see that it is in the desired span without imposing any additional conditions.

Consider now the commutator $[\bar Z_s,\hat Z_t]$. It lies in the span of the
 $\bar Z's$ and the $\hat Z's$ and must be vertical. The only vertical vector in this span is $0$.  So,
$$0 = [\bar Z_s,\hat Z_t] = $$
$$2[r_{stv} - \tilde g_{Jsv,Jt} + \tilde g_{Jsu}\G_{(Jt)v}^u - \tilde g_{tv,s} + \tilde g_{tu}\G_{sv}^u]\frac{\partial}{\partial p_v}
+ $$
$$2i[r_{s(Jt)Jv} + \tilde g_{sv,t} - \tilde g_{Jsu}\G_{tJv}^u - \tilde g_{tv,s} +
\tilde g_{Jtu}\G_{sJv}^u](-)J^v_y\frac{\partial}{\partial p_y}.$$
Thus,
$$0 =  R^{\G }.$$

The vanishing of  $R^{\G}$  seems to be too strong a
 condition. There are various ways to remove it. We could either
 \begin{enumerate}
 \item drop the condition that the barred and unbarred operators satisfy the given commutation
 relations, as in Remark \ref{escape},
 \item impose aspects of the supersymmetry conditions by restricting the space
 of functions rather than only restricting the geometry, or
 \item modify the definition of the $q's$.
 \end{enumerate}
 The second approach has the interesting effect that decreasing the size of the holonomy
 group could increase the number of states in the Hilbert space.

Presumably the choice of one (if any) of the above solutions is dictated by the physics being modeled. When we consider branes, we will not impose the restrictions arising in this subsection as a consequence of the relations between the barred and unbarred operators.

\subsection{Modifying the supercharges}\label{defsusy}
In this subsection, we explore weakening the vanishing curvature condition by modifying our supercharges. We find that modifications of the supercharges may be interpreted as modifications of the geometry ranging from replacing the affine bundle with a neighborhood of the diagonal in $M\times M$ to classical analogs of noncommutative geometries.

Suppose we modify our construction of the $q's$ in the following fashion. We replace the vectorfield $\tilde g_{st}\frac{\partial}{\partial p_t}$ by the field $\Phi(H_s)$, where
$\Phi$ is a smooth injective real linear map onto the vertical vectors. We assume initially that $\Phi(H_s)$ is a perturbation of $\tilde g_{st}\frac{\partial}{\partial p_t}$ in the following sense.
$$\Phi(H_s)(x,p) = \tilde g_{st}\frac{\partial}{\partial p_t} + \Phi^1(H_s)(x,p),$$
where we assume, say, that
$$\Phi^1(H_s)(x,\mu) = 0,$$
and $\Phi^1(H_s)(x,p)$ is analytic in $p$. The previous commutation relations,
$$0=[Z_s,Z_t]=[\hat Z_s,\hat Z_t] = [\bar Z_s,Z_t],$$
 and their conjugates reduce to the following simple relations:
\begin{equation}\label{comrel1}
[H_s,H_t] = [\Psi(H_s),\Psi(H_t)],
\end{equation}
and
\begin{equation}\label{comrel2}
0 = [\Psi(H_s),H_t] + [H_s,\Psi(H_t)],
\end{equation}
where
$$\Psi:= \Phi\circ J.$$

Corresponding to our expansion of $\Phi$,  we seek $\Psi$ of the form
$$\Psi(H_s) = \tilde g_{Jsu}\frac{\partial}{\partial p_u} + \Psi^1(H_s),$$
where $\Psi^1(H_s)$ vanishes at $p=\mu$ and is analytic in $p$.
Fix $\mu = 0$ coordinates. For every integer $n>0$, we write
$$\Psi^1 = \sum_{j=1}^n\psi^j + O(|p|^{n+1}),$$
with $\psi^j$ a homogeneous polynomial of degree $j$ in $p$.  It is easy to solve order by order for the $\psi^j$ (we have done so up to and including quartic order in the case Levi-Civita connection and of vanishing $B$ field); however, we are forced to allow the horizontal lift data, $V_{su}$, also to vary in $p$ if we have to modify $\Psi$ beyond second order in $p$. We illustrate in a special case.
Assume  the $B$ and $\phi$ fields are trivial and $\Gamma$ is Levi Civita.
Then we may take $\psi^1 = 0$ and satisfy (\ref{comrel1}) by setting
$$\psi^2(H_s) = \frac{1}{3}p_yg^{yw}p_nR^{\G n}_{w(Js)u}\frac{\partial}{\partial p_u}.$$
It is perhaps interesting to note that "the lowest energy component" of this term near $p=0$ - its average over spheres - gives the Ricci tensor.
Note that with this choice of $\psi^2$, $\Phi$ may be identified to quadratic order with the metric obtained by pulling back the metric on $M$ via the exponential map (extended to $T^*M$ via the metric identification with $TM$). Thus to lowest order we have replaced our affine geometry with a neighborhood of the diagonal in $M\times M$. If we wish to terminate our deformation at quadratic order, the second condition (\ref{comrel2}) now reduces to
$$0 = p_yp_nR^{\G n}_{uts;Jw}g^{wy}.$$
This condition is slightly weaker than the condition that the metric be locally symmetric and stronger than the (Yang-Mills type) condition that $D^*R^{\G} = 0,$ for an appropriately defined $D$.  It is not clear to us that there is any geometric obstruction to solving the pair of equalities (\ref{comrel1}) and (\ref{comrel2}) order by order if we allow $V_{su}$ to vary in the fiber also. We have seen in (\ref{discq}) that the latter may lead to the introduction of Chern-Simons like terms to the field strength of the $B$ field.

The above choice of $\psi^2$ is not the unique solution of (\ref{comrel1}) to quadratic order.
Viewing $\psi^2\circ J$ as a 1 form, only its exterior derivative enters (\ref{comrel1}) at linear order. Hence we may shift the above choice of $\psi^2$ by any exact form.

Briefly consider now the case of nontrivial $B$ field. We set
$$\psi^1(H_s) = \frac{1}{2}p_y\tilde g^{yJw}r_{wsu}(0)\frac{\partial}{\partial p_u}$$
to solve (\ref{comrel1}) at zero order. Then (\ref{comrel2}) at zero order becomes
$$0 = \tilde g_{Jtu,s} - \tilde g_{Jsu,t} + b_{sv}\tilde g^{vJw}r_{wtu}(0)/2 - b_{tv}\tilde g^{vJw}r_{wsu}(0)/2 + \tilde g_{Jsv}\G^v_{tu} - \tilde g_{Jtv}\G^v_{su}.$$
Cyclically symmetrizing in $s,t,u$ and using (\ref{jtor}), we obtain the deformation of (\ref{tort2}):
$$i(\partial -\db)\omega_{stu} = H_{stu}
- b_{Jsv}e^{-2\phi}g^{vJw}(r_{wJtJu}(0) - r_{wJuJt}(0))/4$$
$$ - b_{Jtv}e^{-2\phi}g^{vJw}(r_{wJuJs}(0) - r_{wJsJu}(0))/4  - b_{Juv}e^{-2\phi}g^{vJw}(r_{wJsJt}(0) - r_{wJtJs}(0))/4.$$

The nonzero $\psi^1$ alters the condition on $\psi^2$.  Write
$$\psi^2(H_s) = \frac{1}{3}p_yg^{yw}p_nR^{\G n}_{w(Js)u}\frac{\partial}{\partial p_u} + \psi^2_1(H_s).$$
Then at first order (\ref{comrel1}) becomes the relation
$$-\frac{1}{4}p_n\tilde g^{nJw}(r_{wsm}(0)\tilde g^{mJp}r_{ptv}(0) -
r_{wtm}(0)\tilde g^{mJp}r_{psv}(0))\frac{\partial}{\partial p_v} =$$
$$ [\tilde g_{Jsv}\frac{\partial}{\partial p_v},\psi^2_1(H_t)]  + [\psi^2(H_s)_1,\tilde g_{Jtv}\frac{\partial}{\partial p_v}].$$

Interpreting the right hand side of this equation as an exterior derivative, we see that the left hand side must be exact. I.e., computing in $g_{ij} = \delta_{ij}$ coordinates, the $B$ field must satisfy the following quadratic Bianchi identity
$$0 = r_{usp}(0)r_{Jptv}(0) + r_{stp}(0)r_{Jpuv}(0) + r_{tup}(0)r_{Jpsv}(0).$$
This restriction could be modified by allowing the $B$ field and associated connection to vary in the fiber. For example, consider the large $B$ field limit. When $b$ is large and $M$ is flat, we set
$$(\tilde g+b)^{-1}_{jk} = \tilde h^{jk} + \tilde \theta^{jk},$$
where $\tilde h^{jk}$ is symmetric and $\tilde\theta^{jk}$ is skew.
Then we can achieve our commutation relations to order $O(\theta)$ by replacing $b(x)$ and $\tilde g(x)$ by
$b((x^s+\theta^{s m}p_m))$ and $\tilde g((x^s+\theta^{s m}p_m))$. This is the classical analog of the noncommutative geometry flat lift considered in deformation quantization (see \cite{Fed}) and in the context of B fields in
\cite[p.12 Section 7]{S1}.
\section{BPS D branes}
In world sheet descriptions of $D$ branes, one associates to the D brane a
subalgebra of the original superconformal algebra, with the subalgebra vanishing
on the brane. In our target space picture, the vanishing of a subalgebra of
operators would correspond to restricting to a subspace of functions on which
the subalgebra vanishes. Such subspaces might be realized, for example, by
functions constant on leaves of a foliation. As we have indicated in section \ref{MDB}, we will consider a more local model with $D$ branes given
 by Lagrangian submanifolds of affine cotangent bundles. Heuristically (if we
 ignore bundle coefficients),  we
 are considering a dual picture with subspaces of functions being replaced by
 quotient spaces of functions; that is functions defined on submanifolds $Z$.
We model the statement that a D brane preserves 1/2 the supersymmetry in
the following fashion. We restrict the operators $q^+,\bar q^-,$ etc. to act on
functions on a subspace $Z$ of $T_B^*M$. In order to define this restriction
we replace the various vector fields arising in the definition of the $q's$ by
their orthogonal projections onto $TZ$. We
denote the restricted operators by a subscript $T$ and the projected vector
fields by a superscript $T$.   We then interpet the
supersymmetry as the assertion that 1/2 of these restricted operators satisfy some vestigial supersymmetry algebra. Thus, for $B$ branes, we require that
$B_T:= q^+_T + \bar q^+_T$ and $B_T^*:= q^-_T + \bar q^-_T$ square to zero. For $A$ branes, we require that  $A_T:= q^-_T + \bar q^+_T$ and $A_T^*:= q^+_T + \bar q^-_T$ square to zero. We further require that the tangential projection respect the pairing between bosons and fermions in the following sense. The pointwise linear span of the operators generated
by the commutators of the tangential supercharges with functions on $Z$ is precisely $n=dim Z$ dimensional. Thus $[B_T,f]$ and $[B_T^*,f]$ for $B$ branes and $[A_T,f]$ and $[A_T^*,f]$ for $A$ branes span an $n$ dimensional space as $f$ runs over arbitrary smooth functions on $Z$. Equivalently, $[B_T,f]$ (respectively $[A_T,f]$) spans an $n/2$ dimensional space as $f$ runs over functions on $Z$. This dimension condition could possibly be weakened, but we will impose it as stated in this note.

We study the operators :

$$q_T^+:= (\gamma^s+iJ\gamma^s)\psi(\nabla_{H_s^T+iJH_s^T}
 + \nabla_{(\tilde g_{st}+iJ\tilde g_{st})(\frac{\partial}{\partial p_t})^T})/2,$$
 
$$q_T^-:= (\gamma^s-iJ\gamma^s)\psi(\nabla_{H_s^T-iJH_s^T}
 + \nabla_{(\tilde g_{st}-iJ\tilde g_{st})(\frac{\partial}{\partial p_t})^{T}})/2,$$
 
$$\bar q_T^+:= (\gamma^s+iJ\gamma^s)\bar\psi(\nabla_{H_s^T+iJH_s^T}
 - \nabla_{(\tilde g_{st}+iJ\tilde g_{st})(\frac{\partial}{\partial p_t})^{T}})/2,$$
 
$$\bar q_T^-:= (\gamma^s-iJ\gamma^s)\bar\psi(\nabla_{H_s^T-iJH_s^T}
 - \nabla_{(\tilde g_{st}-iJ\tilde g_{st})(\frac{\partial}{\partial p_t})^{T}})/2.$$

An interesting question which we have not explored is whether viewing, as in subsection \ref{defsusy}, some subset of the the supersymmetry equations as modifying the operators rather than constraining the geometry leads to deformations of the geometry of the conormal bundle and thus of  $M$ near the brane. Such deformations might reflect the warping of the local geometry by the brane or perhaps recover supergravity brane solutions.
\subsection{Frames adapted to $Z$}
In this subsection we choose a frame for the tangent vectors to $Z$ and fix some
useful notation and formulae.

Pick affine coordinates in the fiber so that we may identify $Z$ with the 
conormal bundle of
a submanifold $Y$ of M. Choose coordinates on the base so that locally our
 submanifold is given by $x^a = 0,$ $a > d+1$. Indices $a,b,c$ will take values
 greater than $d+1$. Indices $i,j,k,l,m,n$ will take values less than or equal
 to $d+1$. Indices $r,s,t,u,v$ will take all values.   
With these conventions, the tangent vectors to $Z$ are given by 
$\frac{\partial}{\partial x^i}$ and $\frac{\partial}{\partial p_a}.$ 

We choose coordinates on the base satisfying
 for all $p$ in $Y$ (intersect the coordinate neighborhood)
$$g_{ia}(p) = 0,$$
for all $i,a$ and 
$$g_{ab}(p) = \delta_{ab}.$$
We choose coordinates in the fiber so that 
$$\mu_a = 0 \mbox{    and   }\mu_{j,a} = 0 = b_{ja}$$
on $Y$. 
We may choose coordinates at a point $p$ so that
$$\G_{ja}^c(p) = 0= V_{ja}(p).$$
We set 
$$X_i = \frac{\partial}{\partial x^i} + V_{ic}\frac{\partial}{\partial p_c}.$$
Then $X_i$ is perpendicular to $\frac{\partial}{\partial p_a},$
 and
$$(\frac{\partial}{\partial p_j},X_i) = -V_{iu}\tilde g^{uj}.$$
Here we have taken the metric in the fiber to be
$$(\frac{\partial}{\partial p_s},\frac{\partial}{\partial p_t}) = \tilde g^{st},$$
 Let
$$\sigma_{ij} = (X_i,X_j).$$
 Then 
$$\sigma_{ij} =  g_{ij} + V_{ik}\tilde g^{ks}V_{js}.$$

With this notation, we have  
$$H_s^T = g_{si}\sigma^{ij}X_j,$$
and 
$$(\frac{\partial}{\partial p_s})^{T} = 
\tilde g^{sb}\tilde g_{bc}\frac{\partial}{\partial p_c} +
(\frac{\partial}{\partial p_s},X_k)\sigma^{kj}X_j.$$
We record some useful commutators. 
$$[X_i,X_j] = (V_{ja,i} - V_{ia,j} + 
V_{ic}V_{ja}^{,c}-V_{ia}^{,c}V_{jc})\frac{\partial}{\partial p_a} .$$
$$[X_i,\frac{\partial}{\partial p_a}] = 
- \G_{ic}^{a}\frac{\partial}{\partial p_c}.$$

The assumption that $Z$ is $\Lambda-$Lagrangian is equivalent to the conditions that  $V_{ij}-\Lambda_{ij}$ is symmetric; thus
$$T^a_{ik} = 0,$$
and
$$b_{ik} = (d\mu)_{ik}/2 + \mu_jT^j_{ik} + \Lambda_{ik}.$$

\subsection{B branes}
In this subsection we show that in our model,  BPS B type branes are given generically by hermitian complex submanifolds equipped with holomorphic bundles.
 We set
$$B_T:= q_T^+ + \bar q_T^+ = $$
$$(\gamma^s+iJ\gamma^s)(\psi+\bar\psi)\nabla_{H_s^T}
+ (\gamma^s+iJ\gamma^s)(\psi-\bar\psi)\nabla_{\tilde g_{st}(\frac{\partial}{\partial p_t})^T} = $$
$$\hat e^j\nabla_{X_j} + f_c\nabla_{\frac{\partial}{\partial p_c}},$$
where
$$\hat e^j = g_{si}\sigma^{ij}(\gamma^s+iJ\gamma^s)(\psi+\bar\psi)  -
V_{ks}\sigma^{kj}(\gamma^s+iJ\gamma^s)(\psi-\bar\psi),$$
and
$$f_c = (\gamma^c+iJ\gamma^c)(\psi-\bar\psi).$$

First consider the supersymmetry condition that $[B_T,f]$ span an $n/2$ dimensional space as $f$ runs
over functions on $Z$. Taking $f$ to run over the coordinate functions, we see that this requires that
the tangent (and normal) directions to $Y$ be stable under $J$. Thus $Y$ is a complex submanifold. In
addition, we see that $V_{ij}$ must satisfy
\begin{equation}\label{herm1}
V_{JiJj} = V_{ij}.
\end{equation}
This implies $\sigma_{ij}$ is also hermitian:
$$\sigma_{JiJj} = \sigma_{ij}.$$
Under the simultaneous assumptions that (\ref{jtor}) holds and that our brane is
Lagrangian in our extended sense and that $\mu$ does not vary in the fiber, (\ref{herm1})  implies that $Y$ is totally
 geodesic, and
 $$\mu_{Ji;Jj} + \mu_{Jj;Ji} = \mu_{i;j} + \mu_{j;i}.$$
(We may remove the totally geodesic condition on $Y$ if we change our choice of connection in the restricted operators or allow $\mu$ to vary in the affine fibers.)

Now consider the supersymmetry requirement
$$B_T^2 = 0.$$
This gives the following separate equations.

$$0 = \hat e^j\hat e^kR(X_j,X_k)/2 + \hat e^jf_cR(X_j,\frac{\partial}{\partial p_c})
 + f_bf_cR(\frac{\partial}{\partial p_b},\frac{\partial}{\partial p_c})/2.$$
\begin{equation}\label{dom}
0 = \hat e^j\hat e^k([X_j,X_k],\frac{\partial}{\partial p_c})/2 + \hat e^jf_b(-V_{jc}^{,b})
+  \hat e^j[\nabla_{X_j},f_c] + f_b[\nabla_{\frac{\partial}{\partial p_b}},f_c].
\end{equation}
$$0 = \hat e^j[\nabla_{X_j},\hat e^k] +
f_b[\nabla_{\frac{\partial}{\partial p_b}},\hat e^k].$$

As our connection splits locally as an orthogonal connection plus the term associated with the $U(1)$ charge, then we may write
$$R = R^{U(1)} + R^0,$$
where $R^0$ is trace free and
$$R^{U(1)}(X_s,X_t) = X_s(iaJ_t^u(p_u-\mu_u)) - X_t(iaJ_s^u(p_u-\mu_u)).$$
Our supersymmetry condition now requires that
$$0 = \hat e^j\hat e^kR^{U(1)}(X_j,X_k).$$
This reduces to the condition that the $(0,1)$ component of $\mu$ is $\db$ closed;
 thus $\mu$ defines a connection on a holomorphic line bundle over $Y$.
 We will see in subsection \ref{singsheaf} how to include more general sheaves in this model.
\subsection{A branes}
In this subsection, we will show that $A$ branes in our model correspond to coisotropic submanifolds of $M$. We will show how to choose fermions so that in the Lagrangian case these are also special Lagrangian.

We define the operator
$$A_T:= q_T^- + \bar q_T^+  = \bar E^s((g_{si}\sigma^{ij}
 - V_{ks}\sigma^{kj})\nabla_{X_j} + \tilde g_{sb}\nabla_{\frac{\partial}{\partial p_b}})$$
$$+ e^s((\tilde g_{si}\sigma^{ij} + V_{ks}\sigma^{kj})\nabla_{X_j} -
\tilde g_{sb}\nabla_{\frac{\partial}{\partial p_b}}),$$
Set
$$\epsilon^j:=  (\bar E^s+e^s)g_{si}\sigma^{ij} - (\bar E^s-e^s)V_{is}\sigma^{ij} : = \epsilon_i\sigma^{ij},$$
and
$$\varphi^b := (\bar E^b -  e^b).$$
Then we may rewrite $A_T$ as
$$A_T = \epsilon^j \nabla_{X_j} + \varphi^a \nabla_{\frac{\partial}{\partial p_a}}.$$

The operators $(\bar E^b-e^b)$ are linearly independent for $b$ distinct. Hence the assumption that the span of the $[A_T,f]$ is $n/2$ dimensional imposes the condition that
$codim Y\leq n/2$.
Fixing a point and coordinates at the point so that $V_{jc}=0$ and $g_{ij} = \delta_{ij}$ we expand
$$\epsilon_j =  \g^j(\psi+\bar\psi) - iJ\g^j(\psi -\bar\psi) - \g^k(\psi-\bar\psi)V_{jk} + iJ\g^k(\psi + \bar\psi)V_{jk},$$
and
$$\varphi^b = \g^b(\psi -\bar\psi) - iJ\g^b(\psi+\bar\psi).$$
From this expansion, it is easy to see that our dimension restriction forces $J_a^b = 0$. In
other words, $J$ maps the normal bundle of $Y$ to its orthogonal
complement. Hence $Y$ must be coisotropic (see \cite{KOA}).

Write
$$\tilde \epsilon_j:= \epsilon_j +iJ^j_a\phi^a =$$
$$ (iJ^j_k+V_{jk})[-\g^k(\psi -\bar\psi) + iJ\g^k(\psi + \bar\psi)].$$
Clearly the $\tilde \epsilon_j$ are linearly independent from the $\varphi's$.
Hence there can be at most $n/2 - codim Y$ real eigenvectors of $(iJ^j_k+V_{jk})$ with
nonzero eigenvalue. This is analogous to the relation between the complex
structure and the curvature of a line bundle discussed in \cite[(10)]{KOA}. However,
the reality of $V$ and $J$ makes this condition stronger than indicated in
\cite{KOA} because $iJ^j_k$ and $V_{jk}$ are linearly independent over the reals.
For example, if $dim Y = n/2$ then this forces the two matrices, $iJ^j_k$ and
$V_{jk}$, to vanish separately. If $\mu$ is required to be constant in the fibers, this forces $Y$ to be Lagrangian with respect to $J$ and totally geodesic, and $\mu$
restricted to $Y$ to satisfy
$$\mu_{i;j} + \mu_{j;i}=0.$$
 In other words, $\mu$ restricted to $Y$ determines a Killing vector when $Y$ is minimal dimension and determines a bundle with curvature given by twice the $B$ field.
  This deviation from the stringy observations is, in part, an artifact of
  the Wick rotation and can be remedied by replacing $\tilde g$ by $i\tilde g$ in the definition of the $q's$. This has the effect of replacing the constraint on the eigenvalues of
$(iJ^j_k+V_{jk})$ by a constraint on the eigenvalues of $(J^j_k+V_{jk})$ as expected, albeit only if we view the operator imposed conditions on the geometry as determining the $\Lambda$ in the holonomic condition rather than supplementing an earlier prescribed condition.

We may recover the special Lagrangian predictions for $A$ branes if instead of taking $\psi$ to be a real fermion, we assume that it is complex and satisfies $J\g^j\psi = i\g^j\psi$.
Once we have introduced these complex fermions, it is easy to see how to bring in Calabi Yau geometry. For example, if we assume that $\psi$ is Majorana-Weyl, then we see that $\psi^*$ can be obtained from $\psi$ by Clifford multiplication by an (n,0) form $\Omega$. If we take $\psi$ to commute with the covariant derivative then $\psi^*$ does also if and only if $\Omega$ is covariant constant, and, therefore, $M$ is Calabi-Yau. Let us next study the implications for $A$ branes of this more constrained system.

Under this alternate formalism with complex $\psi$ and $\tilde g\rightarrow i\tilde g$, we have
$$A_T = \epsilon^j\nabla_{X_j} + i\varphi^a\nabla_{\frac{\partial}{\partial p_a}},$$
 where now
 $$\varphi^a = \gamma^a(\psi^*-\bar\psi),$$
 and
 $$\epsilon^j = i\g^s(\psi^*-\bar\psi)(g_{iJs}-V_{is})\sigma^{ij}.$$
Thus we immediately recover the coisotropy of $Y$ and the relation
$ g_{iJk} -  V_{ik}$ has at most $n/2 - codim Y$ nonzero eigenvalues. (Recall here that
$i,k$ index tangential directions.)
This still leaves us with the condition that $Y$ be totally geodesic. This latter condition is easily removed, however, if we slightly alter our construction by replacing the connection restricted to $Y$ by its projection to the tangent space of $Y$. In other words we pass to the induced connection preserving $TY\subset TM_{|Y}.$ Because we also will be treating the  conormal bundle, we make a corresponding projection of the connection there. This change of connection is not generally compatible with the conditions that $\psi$ and $\psi^*$ be covariant constant (restricted now to $Y$) and that the connection be compatible with the Clifford algebra structure. To see this, consider the case that $Y$ is Lagrangian, and let $c(dv_Y)$ denote Clifford multiplication by the volume form, $dv_Y$, of $Y$. Now as $Y$ is Lagrangian, we have
$$\psi^* = ac(dv_Y)\psi,$$
 for some nonzero scalar function $a$ defined by
 $$\Omega_{|Y} = adv_Y.$$ Assuming our connection is metric compatible, it commutes with $c(dv_Y)$.  Hence $\psi^*$ is no longer covariant constant unless $a$ is constant. Thus we see that if we require $\psi$ and $\psi^*$ to be covariant constant on $Y$ with respect to the induced connection, then $Y$ must be special Lagrangian.
\section{Singular branes}
\subsection{B branes and sheaves}\label{singsheaf}
In this subsection we explore possible relations between our $B$ brane construction and coherent sheaves. Consider, as a basic example, the ideal sheaf ${\mathcal I}_D$
 for a smooth divisor $D$. Let $z_{\alpha}$ be a local defining function for $D$
on the open set $U_{\alpha}$. Then an element of
 ${\mathcal I}_D$ can be written $z_{\alpha}f$, for $f$ a holomorphic function.
The exterior  derivative of $z_{\alpha}f$ is given
$$d(z_{\alpha}f) = z_{\alpha}df + z_{\alpha}\frac{dz_{\alpha}}{z_{\alpha}}f.$$
Consider the brane given,  in some choice of affine
coordinates $p^{\alpha}$ on
$U_{\alpha}\setminus (D\cap U_{\alpha})$ by the
graph of the resulting connection form, the {\em singular} one form $\frac{dz_{\alpha}}{z_{\alpha}}$. We will call branes locally defined as the graph of a singular section {\em singular} branes. Clearly smooth affine transformations cannot transform singular branes to nonsingular branes.
Let $g_{\alpha\beta} = z_{\alpha}/z_{\beta}.$ Let ${\mathcal A}^1$ denote the sheaf of smooth 1 forms.
Then we must have 
$$p^{\alpha} - p^{\beta} = d ln(g_{\alpha\beta}).$$
The Cech cocycle $d ln(g_{\alpha\beta})\in C^1(A^1),$ is a coboundary since 
${\mathcal A}^1$ is a flabby sheaf. Hence, there exist smooth 1 forms 
$s_{\alpha}$ so that $\frac{dz_{\alpha}}{z_{\alpha}} + s_{\alpha}$ patches 
together to give a globally defined submanifold of $T^*M$, and we recognize this $n =
dim M$ brane as a submanifold of the trivial affinization of $T^*M$. It is not, however, equivalent under a smooth affine transformation to the brane given by the zero section because it is singular.
The interpretation of $D$ branes as objects in the derived category of sheaves 
suggests the consideration of smooth homotopy classes of formal linear combinations of
our mechanical $D$ branes. For example, let $\alpha$ and $\beta$ be two smooth 
sections of $T^*M$. Let $[\alpha]$ and $[\beta]$ denote the corresponding
branes. Then  $[\alpha]$ is homotopic to  $[\beta]$  through the branes 
$[(1-t)\alpha + t\beta]$,  $0\leq t\leq 1$.  
On the other hand, returning to the example of the ideal sheaf and the structure sheaf and repeating this construction, we see that
$$[(1-t)(\frac{dz_{\alpha}}{z_{\alpha}} + s_{\alpha})], \mbox{   } 0\leq t\leq
1,$$ is not a smooth homotopy between the zero section brane and the ideal sheaf brane because $(1-t)\frac{dz_{\alpha}}{z_{\alpha}}$ is not smooth for $t<1$.
Instead, perhaps after closing the homotopy relation under some appropriate limit, we have that the ideal sheaf is homotopic to the zero section minus a brane given by the conormal bundle of $D$. The adjunction formula suggests that we identify the sheaf ${\mathcal O_D}$ with the conormal bundle of $D$. Hence this homotopy relation can be viewed as a geometric realization of the following exact sequence of sheaves:
$$0\rightarrow {\mathcal I}_D\rightarrow {\mathcal O}\rightarrow 
{\mathcal O_D} \rightarrow 0.$$

\subsection{Stringy Cohomology}
In this subsection, I speculate about ways in which stringy cohomology might be related to $L_2$ cohomology of B type D branes. In fact, the problem of understanding $B$ fields sufficiently well to see how they might effect $L_2$ cohomology was the initial impetus for this investigation. The remarks and computations here are mostly heuristic; many details remain to be clarified.

 Let $X$ be a singular variety, and
let $\pi:\tilde X\rightarrow X$ be a desingularization.  We assume that in the
complement of the singular set of $X$, the Kahler form of $\tilde\omega$ of 
$\tilde X$ pulls back to $X$ as 
$$\omega = \omega_X + a_i\omega_i,\mbox{   }a_i>0, 1\leq i\leq h,$$
where $\omega_i$ is the curvature of a holomorphic line bundle ${\mathcal L}_i$ 
on $\tilde X$. Assume that these bundles admit a flat connection when restricted
 to 
 $\pi^{-1}X_{smooth}$. Let $A_i$ denote the 1 forms on $X_{smooth}$ given by 
 $$A_i = D_{{\mathcal L}_i} - d_{{\mathcal L}_i},$$
 where the connection $D_{{\mathcal L}_i}$ has curvature $\omega_i$ and 
 $d_{{\mathcal L}_i}$ is the flat connection. Let 
 $$\omega^t = \omega_X + t_i\omega_i,$$ and consider the family of smooth
 varieties $X_t$ with Kahler form $\omega^t$. Then the question we wish to investigate is how to compute on $X_0$, from an analytic perspective, the cohomology of $X_t$, $t>0$. Stringy cohomology (\cite{Zas}) gives a physical solution to this problem; several of the mathematical interpretations of stringy cohomology (see for example \cite{Cr}) give algebraic solutions. Here we seek a de Rham type solution to this problem.

  The cohomology of $X_t$ can be realized as the cohomology of the
   operator $B_T = q_T^+ + \bar q_T^+$ when we identify $X_t$ with the $n=dim X_t$ brane defined by the zero section of $T^*X_t$ equipped with the trivial affine structure. This realization of the cohomology does not adequately reflect the metric data. To encode the metric, we instead consider the family of $n$ branes on $\tilde X$ determined by the metric
 $\omega^t$  as follows. In coordinates where $\mu = 0$, equip $T^*\tilde X$
 with the $B$ field $\omega^t$. For this affine bundle to admit Lagrangian
 sections $\sigma^t$, it is necessary that either we allow torsion or that
 $d\sigma = b$.  Let us consider the torsion free route here. The assumption
 that there exists a primitive for $b$ which is a section of our affine bundle
 implies that, for $b$ nontrivial in cohomology, the affine bundle is
 affine inequivalent to $T^*\tilde X$, even though $b$ is globally
 defined.   
 Now consider a $D$ brane given by such a Lagrangian section. The $B_T$
  cohomology of this brane should be isomorphic to the cohomology of $\tilde X$.
 
 We would like to compare $D$ branes determined by $\omega^t$ on $\tilde X$ to a $D$ brane determined by $\omega_X=\omega^0$ on $X$. In order to do so,
we identify $X_{smooth}$ with $\pi^{-1}(X_{smooth})$. By assumption, 
the $\omega_i$ are exact on $X_{smooth}$, but their primitives may be unbounded.
Nonetheless, the exactness implies that, upon restriction to $X_{smooth}$, 
 there is an unbounded affine equivalence
between the affine structure determined by $\omega^t$ and that determined by
$\omega^0$.  
We see this equivalence as follows. Let $v_a^t$ be primitives for
$\omega^t-\omega^0$ on open sets $U_a$. Hence, $\sigma^t_a-v^t_a$ is a primitive
for $\omega^0$.  Let $s^t$ be a primitive for $\omega^t-\omega^0$ on
$X_{smooth}$. Then $v_a^t-s^t = dl^t_a$, for some $l^t_a$. Hence $v^t_a - v^t_b = dl^t_a-dl^t_b$
is a trivial Cech cocycle. In particular, making the change of coordinates
$p_a\rightarrow p_a - dl^t_a$ gives new coordinate systems whose transition 
functions are determined by the transition functions for $\sigma^t_a-v^t_a$, and
thus by $\omega_X$. So, we have two questions
\begin{enumerate}
\item How does the passage from $\sigma^t$ on $\tilde X$ to $s^t$ over
$X_{smooth}$ alter the cohomology?
\item How does the passage from $s^t$ to $s^0 = 0$ alter the cohomology? 
\end{enumerate}
Our expectation is that the first change does not alter the cohomology whereas
the second does, perhaps under additional hypotheses.
We explore the second transition in an example.

Consider $C^2/Z_2$. Then if $\tilde X$ is the blowup of
 the corresponding cone, we can take $\omega_1$ to be 
 $dd_c ln(|z_1|^2+|z_2|^2).$ So, we might take $s^t = ts$, where
 $$s = d_c ln(|z_1|^2+|z_2|^2) =
 \frac{x_1dy_1-y_1dx_1+x_2dy_2-y_2dx_2}{(|z_1|^2+|z_2|^2)}.$$  
Identifying $(z_1,z_2)$ with the quaternion $q=x_1+iy_1+x_2j+y_2k$, the map 
$s$ becomes, in an appropriate frame, the involution
$$s(q) = i\bar q^{-1}.$$

So, we wish to study the family of sections $ts$ as $t\rightarrow 0$. Outside a
neighborhood of the origin in $C^2/Z^2$, the section converges to the zero
section. In order to understand the behavior interior to a ball of radius
$\epsilon$ about the origin, it is useful to change perspective and view the
section $s^t$ as the graph of a function $\phi^t$ from 
$(T_0^*C^2)\setminus B_{\frac{t}{\epsilon}}(0)$ to the affine
 bundle, given in suitable quaternionic coordinates by
$$\phi^t(p) = it\bar p^{-1}.$$ 
This converges in any reasonable sense to the fiber $T^*_0C^2$, a D0 brane.
So, our limiting brane, as $t\rightarrow 0$ becomes $C^2/Z^2\cup T^*_0C^2$.
What is the $B_T$ cohomology of this
brane? Presumably, it is given by
the sum of the cohomology of the 2 components.  Observe that the vertical
component is contractible, but its cohomology operator $B_T$ is
 given by 
$$(\g^i+iJ\g^i)(\psi-\tilde\psi)\frac{\partial}{\partial \bar w_i}
 + a(\g^i+iJ\g^i)(\psi+\tilde\psi)w_i,$$
$w_i$ complex coordinates on the fiber.  In particular, 
$B_T + B_T^*$ squares to a multiple of
$$\Delta + a^2|z|^2 + ia\g^j(\psi-\bar \psi)J\g^j(\psi+\bar \psi)/2.$$ 
This is a superharmonic oscillator and has a one dimensional kernel.  
Hence, in this example, the limiting brane cohomology is 1 dimensional, as is the stringy cohomology. It would be interesting to prove rigorously the results sketched here and to generalize to wider classes of singularities. It is not clear, however, that this will lead to an effective method for computing on $X_0$ the stringy cohomology because determining the limiting vertical component of the brane may be more formidable than determining the cohomology itself.

\section{Noncommutative deformations}
\subsection{U(1) Yang-Mills}
In this section, we use the interpretation of B fields as connections to construct deformations of supersymmetric abelian Yang-Mills.

 The supercharges of the form
$$\gamma^j\psi(\nabla_{\frac{\partial}{\partial x^j}} + (g_{js} + \mu_{j;s} + b_{js} + p_n\Gamma^n_{js})\nabla_{\frac{\partial}{\partial p_s}}),$$
suggest, for flat space,  a supersymmetry transformation
$$\tilde\delta_{\epsilon} A_s = i\bar \epsilon\g^j(g_{js}+b_{js} + \mu_{j,s})\psi.$$
The infinitesimal gauge transformation associated to a function $T$ which acts as
$$\delta_{T}A_i = T_i,$$
fixes $b$ and replaces $\mu_{j,s}$ by $T_{js}$. The commutator of a gauge and susy transformation gives
$$[\tilde\delta_T,\delta_{\epsilon}]A_i = \bar\epsilon T_{ij}\gamma^j\psi.$$
In order to view this as an operator in our algebra, we would need to allow nonconstant susy parameters. We may remedy this by an operator reordering, redefining our supersymmetry transformation to be
$$\delta_{\epsilon} A_s = i\bar \epsilon\g^j(g_{js}+b_{js})\psi + (i\bar \epsilon\g^j\mu_{j}\psi)_s.$$
With this interpretation, we have
\begin{equation}\label{nonrigid}
[\delta_T,\delta_{\epsilon}]A_s = (i\bar \epsilon\g^jT_{j}\psi)_s = \delta_{i\bar \epsilon\g^jT_{j}\psi}A_s.
\end{equation}

The definition of the supersymmetry transformation on fermions implied by the above supercharge is less obvious. Comparing to the undeformed susy U(1) Yang-Mills,
suggests the association
$$\g^0\g^jg_{sj}\nabla_{\frac{\partial}{\partial p_s}}\rightarrow \g^0\g^jA_{j,0}\rightarrow \g^{0j}F_{j0}\rightarrow \g^{pq}F_{pq}/2,$$
where
$$F=dA.$$
If we view the $b$ field as a component of a generalized metric,
we leave the action on fermions undeformed :
$$\delta_{\epsilon}\psi = - F_{pq}\g^{pq}\epsilon/2,$$
but replace the algebra of gamma matrices with
$$\{\g^p,\g^q\} = (g+b)^{-1}_{pq}.$$
This implies that we should leave the fermions gauge invariant.

We compute the closure of this algebra. Let $\epsilon$ and $f$ be susy parameters. Then
$$[\delta_f,\delta_{\epsilon}]A_s = [(g_{js}+b_{js})(- F_{pq}) - (\mu_{j}F_{pq})_s]i\bar\epsilon\g^j\g^{pq}f/2 - (\epsilon\leftrightarrow f) $$
$$ = 2(F_{qs} - (g+b)^{-1}_{pj}(\mu_{j}F_{pq})_s)i\bar\epsilon\g^{q}f.$$
In the flat space, constant $b$ field case this gives
$$2(A_{s,q} - ((g+b)^{-1}_{pj}\mu_{j}F_{pq}+A_q)_s)i\bar\epsilon\g^{q}f.$$
Thus we have closure on Poincare plus gauge as usual, but the gauge parameter is altered by $- (g+b)^{-1}_{pj}\mu_{j}F_{pq}i\bar\epsilon\g^{q}f.$
Computing the commutator on fermions yields
$$[\delta_f,\delta_{\epsilon}]\psi =
-(i\bar f\g_p\psi)_qF_{pq}\g^{pq}\epsilon - (\epsilon\leftrightarrow f).$$
In the flat space, constant $b$ field case, this gives the usual (zero $B$ field) closure relation, and so after a Fierz manipulation can be reduced to Poincare modulo equations of motion.

We now consider the relation between this algebra and the supersymmetric noncommutative Yang-Mills. See, for example, \cite{SW} for a treatment of supersymmetric noncommutative Yang-Mills. Our point of departure is that gauge parameter redefinitions should be determined by the attempt to make the gauge transformations commute with the supersymmetry transformations.

Write
$$(g + b)^{-1}_{jk} = G^{jk} = h^{jk} + \theta^{jk},$$
where $\theta$ is skew and $h$ is symmetric. Write
$$\g_s = G_{js}\g^j.$$
Consider the change of gauge parameter
$$T\rightarrow  T^{[1]} := T + G^{jk}T_kA_j.$$
Write
$$\delta^{[1]}_T = \delta_{T^{[1]}}.$$
Then we have
$$[\delta^{[1]}_T,\delta_{\epsilon}]A_s = (i\bar \epsilon G^{kj}(T^{[1]}-T)_{j}\g_k\psi)_s
- (G^{jk}T_k(i\bar \epsilon G^{ml}\g_m\mu_{l}\psi)_j)_s = O(b^{-2}).$$

If $S$ is a second gauge parameter then
$$[\delta^{[1]}_T,\delta^{[1]}_S]A_s = \delta_{2\theta^{jk}S_kT_j} + O(b^{-2}).$$

Solving order by order for field and gauge parameter redefinitions which make the gauge transformations commute with the supersymmetry transformations, thus leads (if such redefinitions exist) to a noncommutative deformation of the gauge symmetry. By the uniqueness results of \cite{PSS}, this must be equivalent to the usual supersymmetric Yang-Mills theory.
Hence we see that realizing 2 form potentials as connections leads to
an extremely simple formulation of the commutative (but B field deformed) version of noncommutative Yang-Mills.

\subsection{3 form deformations}
We now consider deformations of supersymmetry algebras associated to a 3 form potential $c_{ijk}$ in flat space. The observations of this subsection were aided by closely related computations in \cite{PSS2}.

We first consider an analog for $\bigwedge^2_CT^*$ of the system of operators $q,\bar q$ defined $T^*_BM$. We set

$$q_2^+ = \g^i\psi\nabla_{(\frac{\partial}{\partial x^i})^h}
 + \g_{ik}\psi \nabla_{\frac{\partial}{\partial p_{ik}}},$$
$$q_2^- = \g^i\psi^*\nabla_{(\frac{\partial}{\partial x^i})^h}
 + \g_{ik}\psi^* \nabla_{\frac{\partial}{\partial p_{ik}}},$$
$$\bar q_2^+ = \g^i\bar\psi\nabla_{(\frac{\partial}{\partial x^i})^h}
 + \g_{ik}\bar\psi \nabla_{\frac{\partial}{\partial p_{ik}}},$$
and
$$\bar q_2^- = \g^i\bar\psi^*\nabla_{(\frac{\partial}{\partial x^i})^h}
 + \g_{ik}\bar\psi^*\nabla_{\frac{\partial}{\partial p_{ik}}},$$
 where $\psi$ and $\bar\psi$ are complex spinors which square to zero in their associated Clifford algebra and are in the $+1$ eigenspace of Clifford multiplication by the volume form.
 (For maximal overlap with the $q^{\pm}, \bar q^{\pm}$, we could further assume that at each point $x$ of $M$, $\psi$ and $\bar \psi$ are highest weight vectors of a maximal torus $T_x\in spin(TM).$ This implies the existence of a maximal rank 2 form, $\omega_T$, defined on $M$.)

Dimensional reduction on a trivial circle fiber of a system with the above operators leads to the consideration of quantum mechanical systems defined on bundles with fibers locally modeled on $T^*M\oplus\bigwedge^2T^*M$ with operators of the form
$$q_3 = \g^i\psi\nabla_{(\frac{\partial}{\partial x^i})^h} \pm \g_{i}\psi \nabla_{\frac{\partial}{\partial p_{i}}}
 + \g_{ik}\psi \nabla_{\frac{\partial}{\partial p_{ik}}}.$$
We have not yet analyzed such systems, except to motivate how to realize the dilaton in the $q's$.

The connection and operators we have now associated to a
3 form potential and our examination of deformations associated to 2 form potentials suggest that the 3 form potential deforms a supersymmetry algebra in the following fashion. Define
$$\delta_{\epsilon}p_{jk} = \bar\epsilon\g_{jk}\psi +
c_{jki}\bar\epsilon\g^i\psi + (b_{ik}\bar\epsilon\g^i\psi)_j + (b_{ji}\bar\epsilon\g^i\psi)_k.$$
Let $T$ be a 1 form gauge parameter
 with
 $$\delta_T\psi = 0,$$
 and
 $$\delta_Tp_{jk} = T_{k,j}-T_{j,k}.$$
Then
 \begin{equation}\label{cdef}[\delta_T,\delta_{\epsilon}]p_{jk} = (dT_{ik}\bar\epsilon\g^i\psi)_j + (dT_{ji}\bar\epsilon\g^i\psi)_k = \delta_{W}p_{jk},
 \end{equation}
 where $W$ is the gauge parameter
$$W_k = dT_{ik}\bar\epsilon\g^i\psi.$$
It is more difficult than in the Yang-Mills case to find a modification of the gauge parameter which removes the gauge transformation on the righthand side of (\ref{cdef}) because of the mismatch in multiplicity of $\gamma$ factors. If we assume that we are acting in a 6 dimensional spacetime and that $c$ is large and self dual, then there is a mechanism for matching the $\gamma$ factors to first order in $c^{-1}$. Set
$$h_{mn} = g_{mn} + c_{Im}c^I_{n},$$
where $I$ runs over skew double indices, and
set
$$\Theta^{ijk} = c^{ij}_lh^{lk}.$$
The selfduality constraint on $c$ implies that $h$ is conformal to $g$. Hence,
$\Theta$ is skew. Consider the parameter redefinition
$$\tilde T_m = T_m + dT_{sm}p_{jk}\Theta^{jks}.$$
Assume for simplicity that $c$ and $g$ are constant.
Then
$$[\delta_{\tilde T},\delta_{\epsilon}]p_{mn} = (d(\tilde T-T)_{in}\bar\epsilon\g^i\psi)_m + (d(\tilde T-T)_{mi}\bar\epsilon\g^i\psi)_n$$
$$ - (dT_{sn}(\Theta^{jks}\bar\epsilon\g_{jk}\psi - g_{il}h^{ls}\bar\epsilon\g^i\psi + (b_{ik}\Theta^{jks}\bar\epsilon\g^i\psi)_j +
(b_{ji}\Theta^{jks}\bar\epsilon\g^i\psi)_k))_m$$
$$ + (dT_{sm}(\Theta^{jks}\bar\epsilon\g_{jk}\psi +
(-h^{ls}g_{il})\bar\epsilon\g^i\psi + (\Theta^{jks}b_{ik}\bar\epsilon\g^i\psi)_j + (\Theta^{jks}b_{ji}\bar\epsilon\g^i\psi)_k))_n
 =$$
$$O(c^{-1}).$$
If $S$ is a second gauge parameter, then we have
$$[\delta_{\tilde S},\delta_{\tilde T}]p_{mn} =
(dT_{sn}dS_{jk}c^{jk}_lh^{ls})_m - (dT_{sm}dS_{jk}c^{jk}_lh^{ls})_n - (S\leftrightarrow T) + O(c^{-2}).$$
 Thus we find the first order variation in the gauge algebra of one forms induced by a three form:
$$[S,T] = \frac{1}{3}\Theta^{jkl}i_ji_ki_ldT\wedge dS,$$
where we have used $i_j$ to denote interior product by $\frac{\partial}{\partial x_j}$. Compare to the Poisson deformation associated to a 2 form $\theta$:
$$[f,F] = \theta^{jk}i_ji_k df\wedge dF.$$

There is not an obvious gauge parameter redefinition which allows us to extend the vanishing of $[\delta_{\tilde T},\delta_{\epsilon}]$ to higher order in $c^{-1}$. We have been imposing the vanishing of this commutator as the condition which selects the deformation of the gauge symmetry. Hence, in order for the representation of the 3 form potential as a connection on an affine bundle to define the deformation of the gauge symmetry to higher order, we must choose a normal form (other than zero) for $[\delta_T,\delta_{\epsilon}]$ (or perhaps deform the susy parameters).  We do not pursue this higher order deformation here.
\section{Acknowledgements}
I would like to thank Paul Aspinwall, Ilarion Melnikov, Sonia Paban, Ronen Plesser, and Savdeep Sethi for helpful discussions. This work was supported in part by NSF grant 0204188.

\end{document}